\documentclass[conference]{IEEEtran}
\IEEEoverridecommandlockouts

\usepackage{cite}
\usepackage{amsmath,amssymb,amsfonts}
\usepackage{algorithmic}
\usepackage[dvipdfmx]{graphicx}
\usepackage{textcomp}
\usepackage{xcolor}
\usepackage{multirow}
\def\BibTeX{{\rm B\kern-.05em{\sc i\kern-.025em b}\kern-.08em
    T\kern-.1667em\lower.7ex\hbox{E}\kern-.125emX}}
\begin{document}

\title{SimBlock: A Blockchain Network Simulator \\
}

\author{
\IEEEauthorblockN{Yusuke Aoki\IEEEauthorrefmark{1}, Kai Otsuki\IEEEauthorrefmark{1}, Takeshi Kaneko\IEEEauthorrefmark{1}, Ryohei Banno\IEEEauthorrefmark{2} and Kazuyuki Shudo\IEEEauthorrefmark{3}}
 \IEEEauthorblockA{Tokyo Institute of Technology \\
2-12-1 Ookayama, Meguro-ku, Tokyo, Japan \\
Email: \IEEEauthorrefmark{1}\{aoki.y.au, ootuki.k.aa, kaneko.t.ay\}@m.titech.ac.jp, \IEEEauthorrefmark{2}banno@computer.org, \IEEEauthorrefmark{3}shudo@is.titech.ac.jp}
}

\maketitle

\begin{abstract}
Blockchain, which is a technology for distributedly managing ledger information over multiple nodes without a centralized system, has elicited increasing attention.
Performing experiments on actual blockchains is difficult because a large number of nodes in wide areas are necessary.
In this study, we developed a blockchain network simulator SimBlock for such experiments.
Unlike the existing simulators, SimBlock can easily change behavior of nodes, so that it enables to investigate the influence of nodes' behavior on blockchains.
We compared some simulation results with the measured values in actual blockchains to demonstrate the validity of this simulator．
Furthermore, to show practical usage, we conducted two experiments which clarify the influence of neighbor node selection algorithms and relay networks on the block propagation time.
The simulator could depict the effects of the two techniques on block propagation time.
The simulator will be publicly available in a few months.

\end{abstract}


\section{Introduction}
Blockchains, which are a core technology of cryptocurrency, are eliciting attention owing to its various application possibilities in numerous aside from cryptocurrency.
Two of the fFures of blockchains are the possibility to manage ledger information without a centralized system even in a group including multiple malicious nodes and the difficulty in tampering previously obtained data.
Due to these features, blockchains are used in many cryptocurrencies, and their applications are extensively investigated.
Accordingly, various research subjects are discussed, such as approval time and scalability.
In conducting the research on these issues, blockchain experiment are often necessary.
However, excluding simple experiments completed at a single node, experiments on blockchains are costly.
In case of preparing a node for the experiment in a public blockchain network built over a wide area, information on the entire network are hard to obtain.
If a private experimental network is constructed, the information on the entire network can be obtained. 
Nevertheless, preparing a large number of nodes is costly, as well as experimental conditions and network configuration cannot be easily changed.

In this study, we developed a blockchain network simulator SimBlock for the research on blockchains.
SimBlock is event-driven, that considering block generation and message transmission/reception as events.
This simulator enables to easily implement the algorithm of the neighbor nodes selection.
Given that  block creation time is calculated from the success probability of block generation, reproducing the mining that requires a large calculation power is unnecessary, and a network that involves many nodes can be simulated.
By modifying the block generation probability, simulating various mining algorithms is possible.

The rest of this paper is organized as follows. 
Section \ref{sec:bc} provides an overview of the blockchain as background knowledge.
Section \ref{sec:simulator} explains and evaluates SimBlock.
Section \ref{sec:utilaization} discusses the performed experiments with SimBlock as the application example.
Finally, Section \ref{sec:summary} elaborates the summary and future work.

\section{Blockchain}
\label{sec:bc}

\begin{figure}[t]
  \begin{center}
    \includegraphics[width=8.35cm]{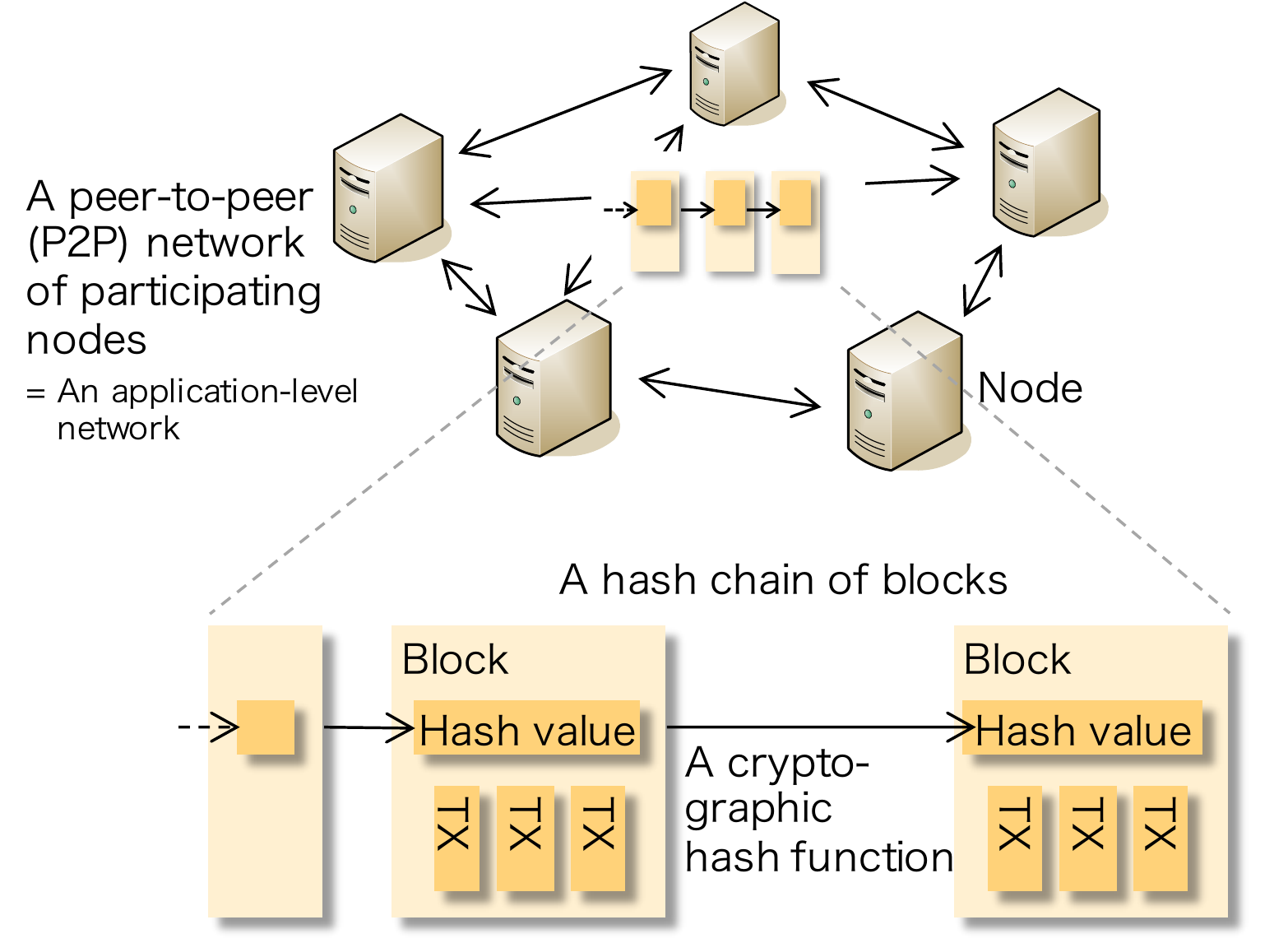}
    \caption{Overview of Blockchain}
    \label{fig:block}
  \end{center}
\end{figure}

In this section, we outline the blockchain as background knowledge.
A blockchain is a distributed ledger technology regarded as part of Bitcoin \cite{nakamoto2008bitcoin}, proposed by Satoshi Nakamoto.
The nodes involved in the blockchain constitute a peer-to-peer network, and a consensus algorithm is established to ensure that the nodes in the network have the same ledger information.
\subsection{Transaction propagation and consensus}

The data to be recorded in blockchains are called transactions and are broadcasted among the nodes involved in a blockchain network.
This transaction is stored in the transaction pool of each node but has not been recorded in the ledger yet.
To store the transaction in the ledger, a block where the plurality of transactions is collected should be generated.
By broadcasting this block, the transactions included in the block are approved and recorded in the ledger.
Each block contains a hash value of the immediately preceding block. Moreover, to change the transactions involved in the past block, rewriting all subsequent blocks is necessary.
Therefore, by adopting a mechanism that appropriately determines the node generating a new block, the blockchains can be stored with difficulty in tampering transactions even in environments with possible multiple malicious nodes.
As shown in Fig. \ref{fig:block}, given that the blocks are connected in a row by including the immediately preceding hash, this technique is called a blockchain.

Several algorithms have been proposed to determine the node that generates the block.
One of the most extensively known algorithms is Proof of Work (PoW) used in Bitcoin.
In PoW, a node that generates a new block is determined based on the computing power of the node.
Each block includes a value freely set by each node called a nonce, and each node locates a block whose hash value of the entire block is below a certain threshold while changing this nonce.
Only the blocks below the threshold are regarded as formal blocks.
Therefore, a node that discovers a nonce satisfying the previously mentioned condition can generate a new block.
The difficulty of the block generation can be adjusted by changing the threshold value.
The process of calculating the hash value of the entire block while changing the nonce is called mining.
In PoW, each node can generate a new block with a probability proportional to the computing power of the node.

\subsection{Network}

The nodes involved in the blockchain form a peer-to-peer network.
Transactions and blocks are broadcasted in this peer-to-peer network.

The nodes participating in the network communicate periodically the obtained information of the nodes.
The node selects a new neighbor node from this node information when a new neighbor node is required.
In Bitcoin's reference implementation Bitcoin Core \cite{Bitcoincore}, new connections are generated at a limited occasion, such as when a node joins the network or when an existing neighbor node is disconnected.
Therefore, the topology of Bitcoin's network does not significantly change in a short period of time \cite{miller2015discovering}.

\begin{figure}[t]
  \begin{center}
    \includegraphics[width=8.35cm]{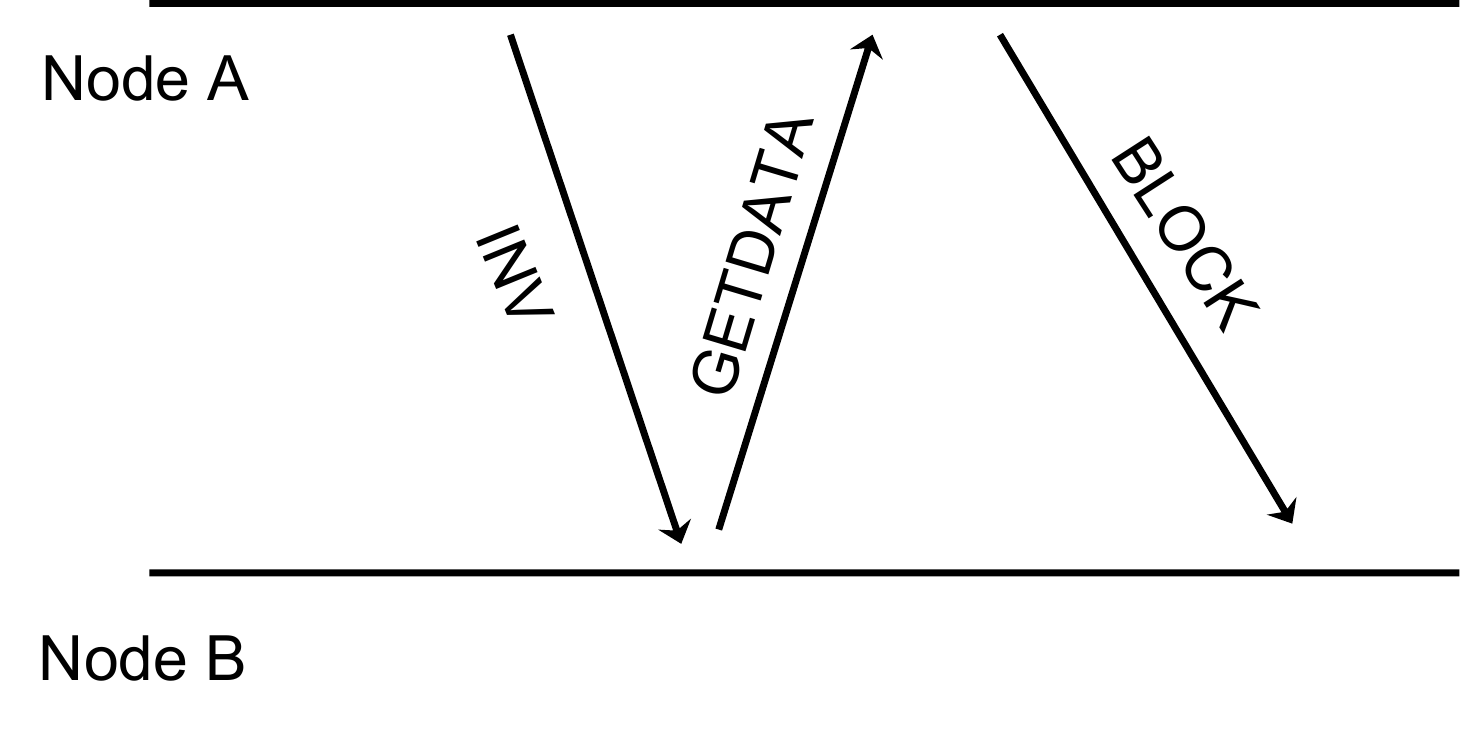}
    \caption{Protocols when Node A sends block to Node B}
    \label{fig:inv}
  \end{center}
\end{figure}

In simple protocols, block transmission/reception is performed using the protocol shown in Fig. \ref{fig:inv}.
If the node that received the INV message does not have the block, then it responds with a GETDATA message and waits for block reception.
By using such protocol, unnecessary transmission of a block with a large data amount does not occur.

\subsection{Fork}
A different block is generated prior to the overall propagation of one block to the entire network; hence, two different types of blocks are propagated on the network, which is called a fork.
When a fork appears, each node has a different block as the latest block, and the data also lose consistency.
To prevent this fork from appearing in existing blockchains, the difficulty of block generation is increased, and the block generation interval is lengthened to ensure that multiple blocks are not simultaneously generated. 
In the case of Bitcoin, the difficulty of block generation is adjusted to wherein one block to generate every 10 min.

\section{Simulator}
\label{sec:simulator}
In this section, we explain the composition of our simulator named SimBlock and evaluate its validity by comparing it with the existing simulator.
\subsection{Design and features}
SimBlock is an event-driven simulator wherein each participating node generates the message and mining events.
In SimBlock, the following values are used as parameters.
\vspace{0.5\baselineskip}
\begin{itemize}
\item \textbf{Block parameters}
	\item[]~\textbf{Block size}: The size of the block generated by the node.
 	\item[]~\textbf{Block generation interval}: Block generation interval targeted by the blockchain.
\setlength{\itemsep}{+0.3cm}
\item \textbf{Node parameters}
\setlength{\itemsep}{0cm}
	\item[]~\textbf{Number of nodes}: The number of nodes involved in the blockchain network. 	
	\item[]~\textbf{Number of neighbor nodes}: The number of neighbor nodes of each node.
	\item[]~\textbf{Location of the node}: The location of each node. Network parameters are determined by the region.
	\item[]~\textbf{Block generation capacity}: The block generation capacity of each node. The block generation difficulty is obtained from the sum of the block generation capacity of all the nodes and the target of the block generation interval. For example in PoW, block generation capacity is a computing power.
\setlength{\itemsep}{+0.3cm}
\item \textbf{Network parameters}
\setlength{\itemsep}{0cm}
	\item[]~\textbf{Network bandwidth}: The upstream and downstream bandwidths for each region. The bandwidth when sending messages from Region A to Region B is regarded as the minimum value of Region A's upstream bandwidth and Region B's downstream bandwidth.
	\item[]~\textbf{Network propagation delay}: The average value of propagation delay between regions. By using this average value, the value according to the Pareto dispersion with a dispersion of 20\% is regarded as the propagation delay.
\end{itemize}
\vspace{0.5\baselineskip}

To calculate the arrival time of a message, we use two parameters as follows: the propagation delay between the nodes and bandwidth.
The transmission time is determined by the message size and bandwidth between the regions, and the message reception event is derived from the total time of the transmission time and propagation delay from the transmission event of the message.
In this simulator, compared with the block message, the other messages are sufficiently small, and the message size is simulated as 0 byte.

When the actual mining is reproduced, the calculation time is high and the number of nodes to be simulated cannot be increased.
Therefore, in the proposed simulator, no algorithm, such as actual hash calculation, is performed.
The time when the mining becomes successful is calculated from the sum of the block generation capacity of all the nodes and block generation difficulty.
In actual blockchains, each node individually determines the block generation difficulty from the generation interval of previous several blocks.
However, considering that the difficulty of all the nodes is the same in usual cases, this simulator uniquely assigns the difficulty to all nodes.
Each node calculates the success time distribution from its block generation capacity and block generation difficulty. 
Further, we simulate the time when the mining becomes successful by deriving a random number that follows that distribution.
Therefore, the identification of the target block generation interval and block generation capacity of each node, it can be applied not only to PoW but also to various consensus algorithms.
In the case of PoW, random number are generated by obtaining the geometric distribution from the difficulty of block generation and the hash rate of the node.

This simulator has the class for managing neighbor nodes.
When a node becomes a target of transmission/reception in each message event or when a node succeeds in generating a block, the corresponding function of the adjacent node management class is called.
Hence, modifying the neighbor node selection algorithm by changing this function becomes possible.

\subsection{Evaluation}
As an evaluation of the simulator, a comparative experiment was performed with the same condition as the existing simulator introduced by Gervais et al. \cite{gervais2016security}.
Their simulator was designed mainly to evaluate its durability against double spoofing attack when modifying the block size and block generation interval.
Therefore, changing the parameters such as block size is easy.
However, altering the consensus algorithm of the node or the algorithm related to the network topology is challenging. Moreover, given that many of the actual blockchain protocols are reproduced, to change the algorithm of the node, adding all the algorithms that correspond to the existing protocol is necessary.

In the experiment, when reproducing the parameters used in the simulator of Gervais et al., we confirm the same simulation result.
We reproduced the actual environment of Bitcoin, Litecoin, and Dogecoin, and evaluated the occurrence rate of the fork and time until the generated block reaches half of the nodes involved in the network.
The parameters in Table \ref{table:actualdata} were reproduced similar to that of Gervais et al. 's simulator.
\begin{table}[t]
\begin{center}
	\caption{Parameters of Gervais et al.'s simulator}
	\label{table:actualdata}
	\begin{tabular} {lccc} \hline
		 Parameter	& Bitcoin 	& Litecoin	&Dogecoin\\ \hline
		\# of the nodes	&6000 	 &800    &600\\
		Block interval	&10 min	&2 min 30 sec	&1 min\\
		Block size	&534 KiB	&6.11 KiB	&8 KiB\\
		\# of the connection &\multicolumn{3}{c}{
Distribution according to Miller et al.  \cite{miller2015discovering}}\\
		Geographical distribution	&\multicolumn{3}{c}{Distribution according to actual blockchains}\\
		Bandwidth	 & \multicolumn{3}{c}{\multirow{2}{*}{6 regional bandwidth and propagation delay}}\\
		propagation delay &\\ \hline
	\end{tabular}
	\end{center}
\end{table}
The number of nodes, block size, and distribution area of the nodes is the result of Gervais et al.'s observation of the actual blockchain in 2015.
We set up six regions, namely, Europe, North America, Asia, Australia, Japan, and South America, and the bandwidth and propagation delay of each region were set by reproducing the network at the time.
The nodes were distributed in these six regions based on the actual data, and the bandwidth and propagation delay between the regions were applied to the nodes.
The propagation delay followed a Pareto distribution with a mean value based on actual data and variance of 20\%.
The number of connection distribution of the nodes is based on the observation of Miller et al.  \cite{miller2015discovering}.
Each node randomly selected the nodes from the entire network and set it as its own neighbor nodes.
However, the distribution of the nodes' block generation capacity in the actual environment cannot be measured.
In this experiment, the normal distribution with a standard deviation of one-third of the average was regarded as the distribution of the block generation capacity.

We simulated it until 10,000 blocks were generated.
The result of SimBlock and that of Gervais et al. and real data is shown in Table \ref{table:CCS2016}.
\begin{table}[t]
\begin{center}
\caption{Median block propagation time ($t_{MBP}$) and rate of the fork ($r_f$) in actual networks and simulator. }
	\label{table:CCS2016}
	\begin{tabular} {lccc} \hline
				     & Bitcoin 	 & Litecoin & Dogecoin\\ \hline
		Block interval	     &10 min 	 & 2.5 min   & 1 min\\\hline
		Measured $t_{MBP}$ & 8.7 s	 & 1.02 s	   &  0.98 s\\
		Gervais et al. $t_{MBP}$ & 9.42 s & 0.86 s & 0.83 s\\
		SimBlock $t_{MBP}$ & 8.94 s & 0.85 s & 0.82 s\\\hline
		Measured $r_f$ 	& 0.41 \% & 0.27 \%& 0.62 \% \\
		Gervais et al. $r_f$&1.85 \% & 0.24 \% &0.79 \% \\
		SimBlock $r_f$ &0.58 \%&0.30 \%&0.80 \% \\ \hline
	\end{tabular}
	\end{center}
\end{table}

Every result is close to the measured value and that of Gervais et al., and the proposed simulator can simulate the blockchains with good accuracy.
Specifically, SimBlock can simulate values that are very close to that of Gervais et al. who used the same network parameters.
Thus, a more accurate simulation of the actual value is possible by reviewing the parameters.
Only the rate of the fork in Bitcoin obtained a large error because  Bitcoin uses a relay network that is not reproduced in the experiment.

\section{application example of the simulator}
\label{sec:utilaization}
For the application examples of the simulator, we performed an experiment where the neighbor node selection algorithm and experiment in which the participation rate in the relay network were modified.

\subsection{Purpose}
One of the problems of existing blockchains is low transaction throughput.
The throughput of the blockchain is the quotient obtained by dividing the number of transactions included in the block by the block generation interval.
In the case of Bitcoin, the upper limit of the number of transactions included in one block is approximately 4000, and the block generation interval is 10 min. Therefore, the upper limit of the throughput is approximately seven transactions per second.
This throughput is very small compared with approximately 1700 transactions per second \cite{Visa}, which is the average throughput of Visa, and approximately 290 transactions per second, which is the average throughput of PayPal \cite{Paypal}.
One technique to address this problem is reducing the block propagation time.
By shortening the propagation delay, we can safely reduce the block generation interval and improve the throughput \cite{decker2013information}.
We provide two examples of methods that can reduce the block propagation time and simulate blockchains by adopting these methods.

The first method is attempting to improve the efficiency of the network topology.
As mentioned in Section \ref{sec:bc}, a blockchain network is peer-to-peer without a central administrator.
Therefore, the topology of the network is dependent on the manner of selecting the neighbor nodes of individual nodes.

The second one is preparing a block propagation-dedicated network different from the network used in the blockchain.
As existing research, bloXroute \cite {klarmanbloxroute} and Falcon \cite {Falcon} proposed a block propagation-dedicated network and attempted to improve the block propagation efficiency.
To measure the effect of such a relay network, we observe the propagation time while changing the participation rate of the node to the relay network on the simulator.

\subsection{Algorithm of the neighbor node selection}
In the proposed algorithm,  we designed each node to connect with the node that sent the INV message earlier to it.
Each node scores the node that sent the INV message to it and determines the connection priority.
Each time an INV message is received, each node records the elapsed time since the block creation time of the INV message.
Each time a node receives 10 blocks, each node updates the neighbor node based on the average of the recorded elapsed time.
For each node sent the INV message, sets the weighted average of the recorded elapsed time as a score.
Up to the maximum number of connections, connections are established as a new neighbor node from the smallest to the higher score.
However, to obtain information on the new node, one neighbor node is randomly selected from all the nodes.

\begin{figure}[t]
  \begin{center}
    \includegraphics[width=8.35cm]{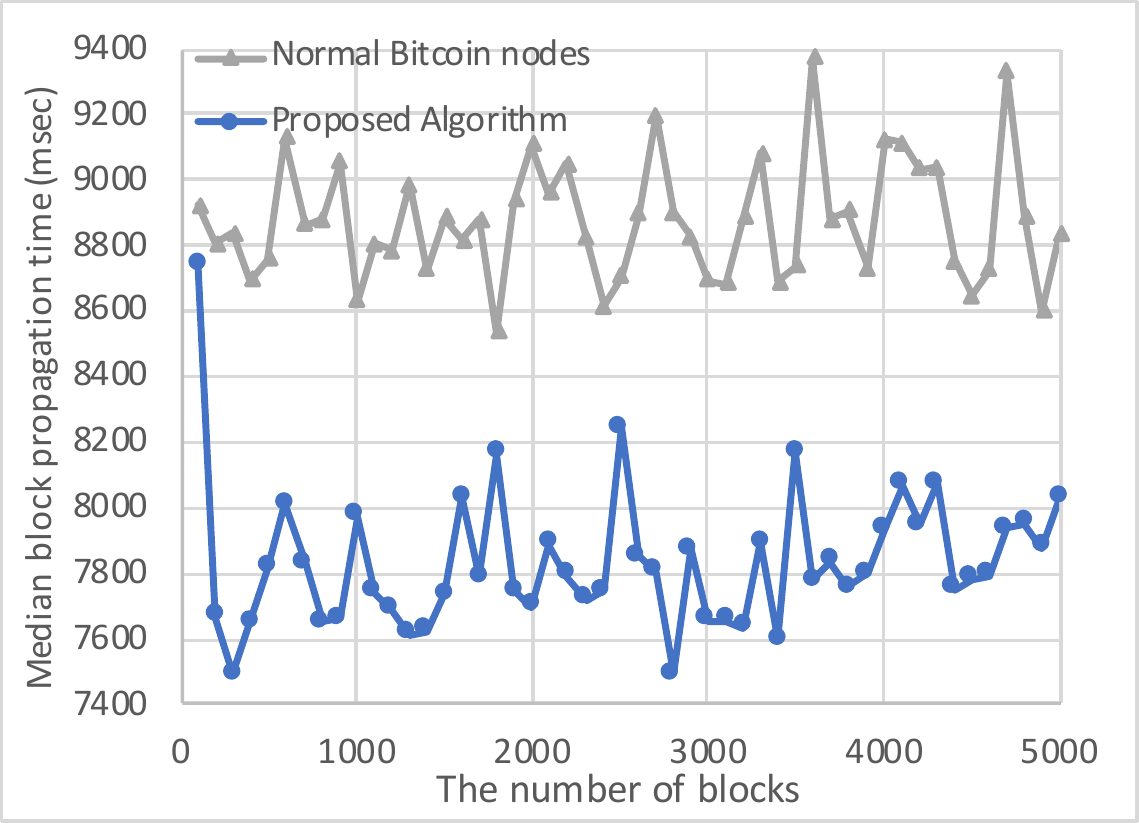}
    \caption{Median of the block propagation time}
    \label{fig:result}
  \end{center}
\end{figure}
We performed the experiments by using the same parameters that of Bitcoin, as discussed in Section \ref{sec:simulator}.
We compared the median value of the block propagation time in networks composed of nodes with fixed neighbor nodes and a network consisting only of nodes adopting the proposed node selection algorithm.
The former reproduces the ordinary Bitcoin node.
The graph is shown in Fig. \ref{fig:result} where the horizontal axis denotes the number of generated blocks and the vertical axis indicates the median of the block propagation time.
The median of the block propagation time is averaged every 100 blocks and then plotted.

The experimental result shows that the block propagation time of the entire network is improved using the proposed neighbor node selection algorithm.
It also demonstrates that the propagation time is sufficiently improved in a small number of times by replacing the neighbor nodes.
However, no improvement in the propagation time is further observed after 100 blocks are generated.
In the proposed algorithm of this experiment, the number randomly selected as an adjacent node from all the nodes is set to 1.
However, by changing this number, the improved speed and limit of the propagation time may change.

As observed in the figure, the propagation time in the nodes with smaller calculation power is larger than the propagation time in all the nodes.
This observation seems to be because the nodes with large calculation power transitions to the center of the network as the result of the neighbor node selection.
Although we could prove that the proposed algorithm can improve the block propagation time, we will also investigate the effect of such network bias on the security and benefits of individual nodes in the future.

\subsection{Relay network}
Constructing a relay network for block distribution is proposed as the method used in reducing the block propagation time in blockchains.
The block propagation mechanism of the blockchains operates even in an environment where malicious nodes exist, but it is not always suitable when focusing on propagation efficiency.
The relay network is a network for block propagation developed outside the mechanism of blockchains, and sending and receiving blocks among the participating nodes earlier than the normal block propagation in blockchains are possible.

The mechanism for improving the efficiency of the block propagation in a relay network varies depending on the implementation, but we conceptualized this mechanism in the experiment.
The nodes involved in the relay network are assumed to transmit the block by using the usual 10 times the bandwidth to the other nodes participating in the relay network.
While changing the proportion of the nodes involved in the relay network, the median value of the block propagation time was measured.
The median value of the propagation time was measured for the three types of node groups, namely, all the nodes in the blockchain, nodes involved in the relay network, and nodes not involved in the relay network.
The parameters were set under the same conditions as that of Bitcoin, as discussed in Section \ref{sec:simulator}.
\begin{figure}[t]
  \begin{center}
    \includegraphics[width=8.35cm]{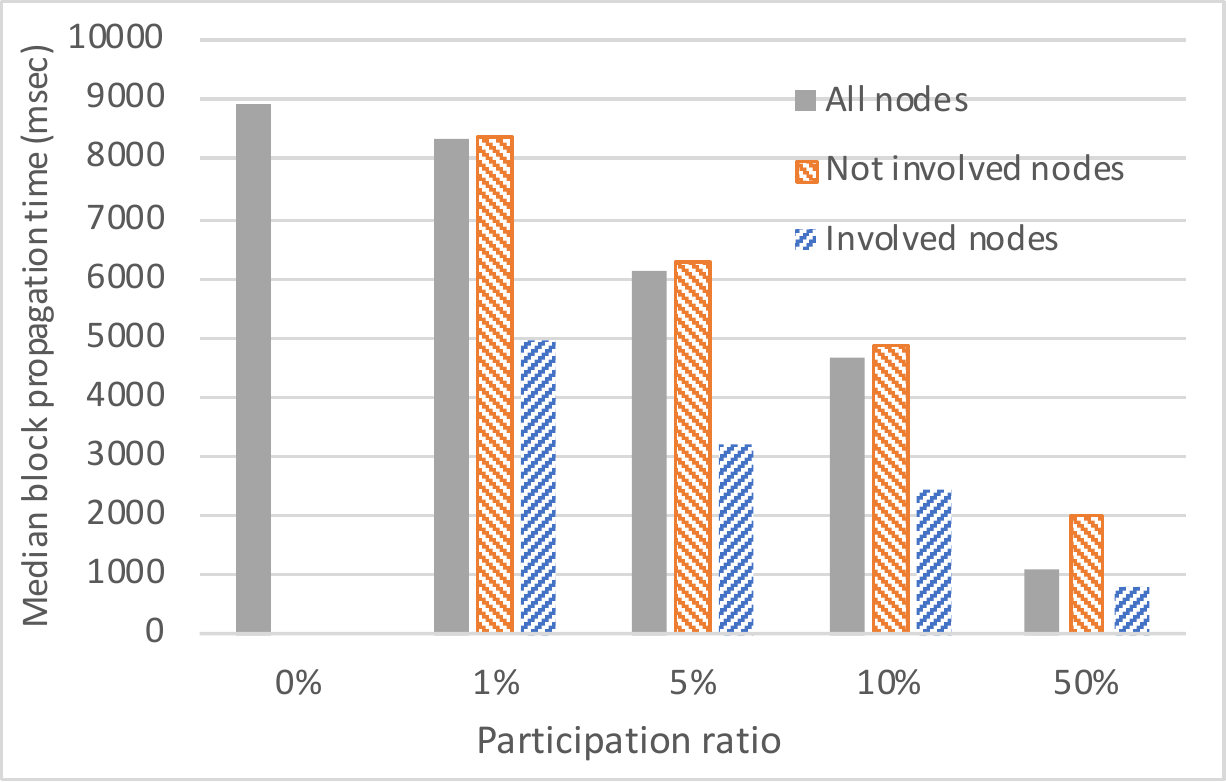}
    \caption{Median of the block propagation time}
    \label{fig:result3}
  \end{center}
\end{figure}

Fig. \ref{fig:result3} shows the simulation results.
Even if the participation rate in the relay network is as low as 5\%, the propagation time of all nodes is remarkably improved to less than 70\% of the original propagation time.
The propagation time among the nodes participating in the relay network is further improved remarkably.
As the involved rate in the relay network increases, the difference in the propagation efficiency between the non-participating and participating nodes increases.

\section{Summary and future work}
\label{sec:summary}
In this study, we proposed a blockchain simulator named as SimBlock.
We confirmed that the simulator could simulate an actual blockchain with good accuracy.
We also presented the techniques in using the simulator and demonstrated that the simulator is useful in research.

Future work includes further expansion of the simulator.
Although the current simulator simulates a simple block transmission protocol, we plan to add an implementation that also supports the latest transmission protocol, such as the compact block.
We also want to include a simulation of transactions that have not been conducted this time. 
We believe that some mechanisms used for sending and receiving blocks can be partially added by sending and receiving transactions.

In a few months, the simulator will be released on the website after setting up the input method of the experimental scenario and data output method.

\section*{Acknowledgment}
This work was supported by JSPS KAKENHI Grant Numbers 16K12406, New Energy and Industrial Technology
Development Organization (NEDO), and SECOM Science and Technology Foundation.

\bibliography{bib}

\begin{thebibliography}{1}

\bibitem{nakamoto2008bitcoin}
Satoshi Nakamoto.
\newblock Bitcoin: A peer-to-peer electronic cash system.
\newblock 2008.

\bibitem{Bitcoincore}
{Bitcoin Core : Bitcoin}.
\newblock https://bitcoincore.org, (accessed Jan. 10, 2019).

\bibitem{miller2015discovering}
Andrew Miller, James Litton, Andrew Pachulski, Neal Gupta, Dave Levin, Neil
  Spring, and Bobby Bhattacharjee.
\newblock Discovering bitcoin’s public topology and influential nodes.
\newblock 2015.

\bibitem{gervais2016security}
Arthur Gervais, Ghassan~O Karame, Karl W{\"u}st, Vasileios Glykantzis, Hubert
  Ritzdorf, and Srdjan Capkun.
\newblock On the security and performance of proof of work blockchains.
\newblock In {\em Proceedings of the 2016 ACM SIGSAC Conference on Computer and
  Communications Security}, pages 3--16. ACM, 2016.

\bibitem{Visa}
{VisaNet | Electronic Payments Network}.
\newblock https://usa.visa.com/about-visa/visanet.html, (accessed Jan. 10,
  2019).

\bibitem{Paypal}
{PayPal, Inc | PayPal Reports Second Quarter 2018 Results}.
\newblock
  https://investor.paypal-corp.com/news-releases/news-release-details/paypal-reports-second-quarter-2018-results?ReleaseID=1072972,
  (accessed Jan. 10, 2019).

\bibitem{decker2013information}
Christian Decker and Roger Wattenhofer.
\newblock Information propagation in the bitcoin network.
\newblock In {\em Peer-to-Peer Computing (P2P), 2013 IEEE Thirteenth
  International Conference on}, pages 1--10. IEEE, 2013.

\bibitem{klarmanbloxroute}
Uri Klarman, Soumya Basu, Aleksandar Kuzmanovic, and Emin~G{\"u}n Sirer.
\newblock blo{X}route: A scalable trustless blockchain distribution network
  whitepaper.

\bibitem{Falcon}
{Falcon - A Fast Bitcoin Backbone}.
\newblock https://www.falcon-net.org/, (accessed Jan. 10, 2019).

\end{thebibliography}
\bibliographystyle{unsrt}

\vspace{12pt}

\end{document}